\documentclass[fleqn,twoside]{article}
\usepackage{espcrc2}
\usepackage{graphicx}
\usepackage{url}
\usepackage{float}
\restylefloat{figure}
\usepackage[figuresright]{rotating}

\newcommand{\AmS}{{\protect\the\textfont2
  A\kern-.1667em\lower.5ex\hbox{M}\kern-.125emS}}

\title{A perturbative determination of the parameters of an anisotropic 
quark action 
}
\author{Justin Foley\address{
The TrinLat Collaboration,
School of Mathematics, Trinity College, Dublin 2, Ireland}
        \thanks{Talk presented by Justin Foley}, 
	Mike Peardon\addressmark \hspace{0.5mm}, Sin\'{e}ad M. Ryan\addressmark}
\begin{document}

\begin{abstract}
The parameters of a 3+1 anisotropic quark action
with Symanzik-improved glue are determined at 1-loop in
perturbation theory. 
\end{abstract}

\maketitle
 
\section{INTRODUCTION}
The utility of the anisotropic lattice formulation in glueball 
spectroscopy~\cite{gb} is well documented. 
In that case, using a lattice spacing which is finer in 
the temporal direction, $a_{t}$, than in the spatial, $a_{s}$,
gives 
improved resolution of correlation-function decays
with a minimal increase in computational workload. 
Our recent work has centred on the application 
of anisotropic lattices to the 
simulation of relativistic heavy quarks.  
We use a tree-level improved 
fermionic action, specifically formulated 
for large anisotropies on a 3+1 lattice, which 
is free of classical discretisation errors 
that scale with  $ a_{s} m_{q} $. 
If the cutoff effects remain well-defined beyond the tree level,
 it should be possible to accurately 
simulate heavy-quark systems on lattices 
for which $ a_{s}\gg a_{t} $.

\subsection{ Anisotropy }

Of critical importance is the dependence of
the renormalised anisotropy $\xi_{R} = \frac{a_{s}} {a_{t}}$,  
determined from simulations, on both the bare anisotropy, $\xi$, and 
the bare quark mass. This report describes the tuning of the 
quark action parameters required to restore Lorentz symmetry 
(i.e. to restore the renormalised speed of light, $c_{R} \equiv \xi_{R}^{-1} \xi$, to 
its correct physical value of unity) at 1-loop in perturbation theory. 
An analytic study 
of this action in the Hamiltonian limit,
 $ a_{t} \rightarrow 0 $,~\cite{ho} has yielded promising results. 
In that study, corrections to the speed of light parameter in the action 
were calculated at 1-loop order and 
found to depend almost linearly on the combination $a_{s} m_{q}$. 
The results of our numerical investigation have also been
positive~\cite{aniso}. The variation of the renormalised anisotropy measured 
from the dispersion relations of pseudoscalar and vector mesons 
was shown to be mild over a range of quark masses from strange to charm.
The success of that exploratory work has motivated this more 
comprehensive investigation of the quantum corrections to 
the fermionic action using perturbation 
theory. 

\subsection{Quark Action}
The quark action keeps the usual Wilson term in the temporal direction while using a
Hamber-Wu-like term to remove doublers in the spatial directions
\begin{eqnarray}
 S_{q} & = & \bar{\psi} (\gamma_{0} \nabla_{0} 
+ \sum_{i} \mu_{r} \gamma_{i} 
\nabla_{i}(1 - \frac{1} {6} a^{2}_{s} \triangle_{i}) \nonumber
 \\
& &-  \frac{r a_{t}} {2}
( \triangle_{0} - \frac{1} {2} \sigma_{i 0} F_{i 0} ) 
+ s a^{2}_{s} \sum_{i} \triangle^{2}_{i}
\nonumber \\
& & +  m_{0} ) \psi,
\end{eqnarray}
where $\mu_{r} = 1 + \frac{1} {2} r a_{t} m_{0} $, 
and $r$ and $s$ are constants which we set to 
$1$ and $0.125$ respectively. 
Tadpole improvement factors are contained in the definition of  
the lattice derivatives and the chromoelectric field.    
This action is classically improved to ${\cal O}(a_{t}, a_{s}^{3}) $ and 
further details of its construction can be found in Ref.~\cite{aniso}.

\subsection{Gauge Action}
We use the Symanzik-improved gauge action introduced in Ref.~\cite{gb}
\begin{eqnarray}
S_{w} &=& \frac{\beta} {\xi} \left \{  \frac{5(1 + \omega)} {3 u^{4}_{s}} \Omega_{s} 
- \frac{5 \omega} {3 u^{8}_{s}} \Omega_{s}^{2t} - \frac{1} {12 u^{6}_{s} } \Omega_{s}^{R} \right \} \nonumber \\
& &+ \beta \xi \left \{ \frac{ 4 } { 3 u_{s}^{2} u_{t}^{2} } \Omega_{t}
	- \frac{1} {12 u_{s}^{4} u_{t}^{2} } \Omega_{t}^{R}  \right \},
\end{eqnarray} 
 where $\Omega_{s}$ and $ \Omega_{t}$ are spatial and temporal plaquettes
and $ \Omega_{s}^{R} $ and $ \Omega_{t}^{R} $ are $2 \times 1$ rectangles in the $(i,j)$ and $(i,t)$ planes respectively. 
$ \Omega^{2t}_{s} $ is constructed from two spatial plaquettes separated by a single temporal link.
In simulations, $\omega$ is chosen to avoid a critical point in the fundamental-adjoint action plane 
of $SU(3)$. In perturbation theory, $ \omega $ appears only through
 vertices 
involving four or more gluons and its value
is irrelevant to our calculation.

\section{TUNING}
To calculate the corrections to the 
fermionic action, we consider 
an expansion of the quark energy 
in powers of the spatial momentum 
\begin{eqnarray}
\hspace{1pc}
E^{2}(\mathbf{p}) = M_{1}^{2} + \frac{M_{1}} { M_{2}} \mathbf{p}^{2} + O(p^{4}),
\end{eqnarray}
where $M_{1}$ is the rest mass and $M_{2}$ is the kinetic mass.
Setting $ M_{1} = M_{2} $ restores Lorentz invariance. 
This is the mass-dependent improvement condition suggested in Ref.~\cite{massive}.
The deviation of the measured speed of light from unity can be attributed to two factors. 
Firstly, at non-zero quark mass, the speed of light is modified by lattice artifacts
even in the isotropic 
formalism.
Secondly, in this calculation, additional radiative corrections arise due to the 
different lattice spacing in the spatial and temporal directions and the inherent asymmetry of the 
actions used.
 
The relationship between $M_{1}$, $M_{2}$ and the action parameters
may be determined to any order in perturbation theory 
by expanding the all-orders dispersion relation~\cite{se},
 corresponding to a 
pole in the momentum-space full quark propagator, in powers 
of the coupling.
At the tree level, tuning the action parameters
to satisfy the improvement condition amounts to 
a redefinition of $\mu_{r}$, 
\begin{eqnarray}
\mu_{r}^{(0)} &= & \sqrt{ \frac{ a_{t} m_{0}( 2 + a_{t} m_{0})} 
{2 \ln(1+ a_{t}m_{0})}}, 
\end{eqnarray}
which leaves the action unchanged at ${\cal O}(a_{t})$. Similarly, it is clear from the 
form of the action and the quark dispersion relation that higher-order radiative corrections 
can be also be absorbed into $ \mu_{r}$.
The expression for the 1-loop coefficient is 
\begin{eqnarray} \label{1loop}
\mu_{r}^{(1)} = \Big(\frac{e^{2 a_{t} M_{1}^{(0)}}} { \mu_{r}^{(0)}}  - \mu_{r}^{(0)}
 \Big) 
\frac{M_{1}^{(1)}} {2 M_{1}^{(0)}}
+ \frac{\mu_{r}^{(0)}} {2} Z_{M_{2}}^{(1)},
\end{eqnarray}
where the kinetic mass renormalisation factor,
\begin{eqnarray}
Z_{M_{2}} = \frac{\mu_{r}^{2} M_{2}} { \sinh(a_{t} M_{1}) } e^{-a_{t} M_{1}},	
\end{eqnarray}
gives a measure of the difference
between $M_{1}$ and $M_{2}$ due to radiative corrections.
At 1-loop order, $M_{1}^{(1)}$ and $ Z_{M_{2}}^{(1)} $
are independent of $ \mu_{r}^{(1)} $ and Eq.~\ref{1loop} is a closed expression 
for the 1-loop correction to the action.
The 1-loop coefficients of $M_{1}$ and $Z_{M_{2}}$ have been calculated for a number of different 
quark actions in Refs.~\cite{ho} and~\cite{se}. As these are physical quantities, each term
in their perturbative expansion must be
infrared finite and gauge-invariant. Eq.~\ref{1loop} makes explicit the gauge invariance of $ \mu_{r}^{(1)} $
which serves as an important check in our calculation.

In the limit $ m_{0} \rightarrow 0$, the 1-loop coefficient reduces to the expression used in
Ref.~\cite{groote}.  

\section{DETAILS OF THE CALCULATION}
The calculation of $\mu^{(1)}_{r}$
reduces to a determination of the 1-loop quark self-energy and its derivatives 
with respect to external momenta. Details can be found in Refs.~\cite{ho,se}.
Expressions for the propagators and vertices were obtained using 
Mathematica. The spin algebra required to obtain the loop integrands was initially performed 
with Mathematica but later checked using an automated approach.
To find derivatives,
automatic differentiation~\cite{fad} was used where analytic expressions 
would have been 
too unwieldy.
We used the mean link in Landau gauge to implement tadpole improvement.
Loop integrals were evaluated with Vegas~\cite{vegas}. Although the 1-loop coefficient
is infrared finite, this depends on the cancellation of divergences in its constituent 
pieces. To calculate the integrals numerically,
we introduced a small gluon mass, $\lambda$, as an intermediate regulator.
For a given quark mass and bare anisotropy, we calculated $\mu^{(1)}_{r}$ over a range 
of gluon mass values and extrapolated to $\lambda = 0$ to obtain a final result.

\section{PRELIMINARY RESULTS}
We present results for $\xi = 6$ which was the bare anisotropy 
examined in Ref.~\cite{aniso}. The 1-loop values given below are coefficients of $g^{2}$.  
In this initial study we did not include a chromoelectric term in the 
fermionic action. 

For Wilson-type quarks there is an additive renormalisation of the bare quark mass. We find that at 
$\xi=6$, the 1-loop critical bare mass 
corresponding to massless quarks is given by $a_{t}m^{(1)}_{c} = -0.008688(1)$. 

\begin{figure}[ht]
\begin{center}
\includegraphics[width=70mm]{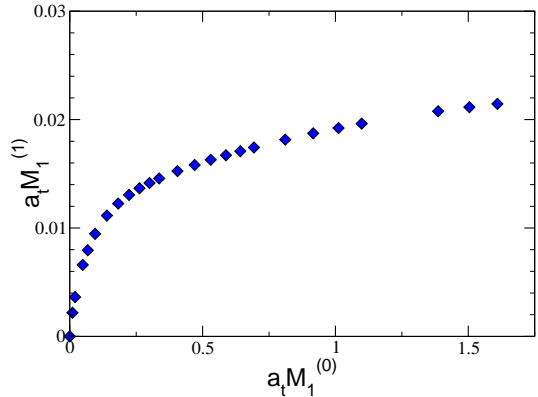}
\caption{The 1-loop rest mass renormalisation at $\xi=6$.}
\vspace{-3ex}
\label{rmass}
\end{center}
\end{figure}

\begin{figure}[ht]
\begin{center}
\includegraphics[width = 70mm]{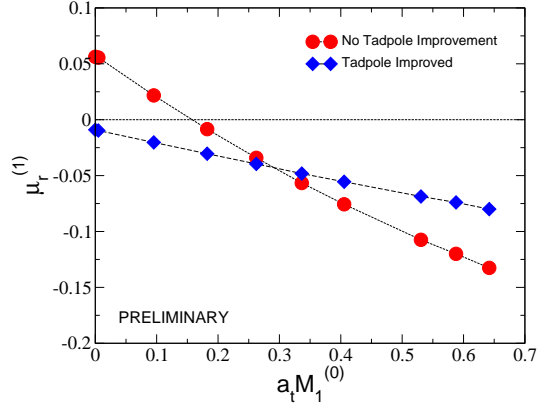}
\caption{Speed of light renormalisation plotted as a function of the tree-level rest mass at $\xi=6$.}
\vspace{-3ex}
\label{sol}
\end{center}
\end{figure}

Fig.~\ref{rmass} shows the 1-loop rest mass as a function of the 
tree-level rest mass. 
$a_{t} M_{1}^{(1)}$ varies smoothly 
and remains small over a range of values of $a_{t} M_{1}^{(0)}$.
The main result, however, is the mass dependence of $\mu^{(1)}_{r}$ which is plotted 
in Fig.~\ref{sol}. The 1-loop correction to $\mu_{r}$ is extremely small and 
appears to vary linearly with the mass. This dependence is very weak and the value of $\mu^{(1)}_{r}$
changes by less than $0.1$ over the  range considered.

\section*{OUTLOOK AND DEVELOPMENTS} 
Preliminary results for the speed of light renormalisation are in qualitative agreement with our previous work.  
We plan to carry out a detailed comparison of perturbative and non-perturbative tuning of the fermionic action in the near future. 
Ultimately, this work will extend to a Symanzik-type perturbative improvement program for the quark action including chromomagnetic 
interactions.

\section*{ACKNOWLEDGMENTS} 
We thank Giuseppe Burgio, Alessandra Feo, Ron Horgan and Quentin Mason for their advice.
J.~F. is supported by an IRCSET postgraduate scholarship.

\end{document}